\documentclass[prl, reprint, amsmath,amssymb,superscriptaddress,nofootinbib,floatfix, aps, longbibliography]{revtex4-1}
\usepackage{graphicx}
\usepackage{dcolumn}
\usepackage{bm}
\usepackage{booktabs}
\usepackage{colortbl}
\usepackage{enumerate}
\usepackage{verbatim}
\usepackage{multirow}
\usepackage{xparse}
\usepackage{tabularx}
\usepackage[colorlinks=true,urlcolor=blue,anchorcolor=black,citecolor=blue,filecolor=black,linkcolor=blue,menucolor=blue,linktocpage=true]{hyperref} 

\begin{document}

\title{Collapse versus Disruption: The Fate of Compact Stellar Systems in Ultralight Dark Matter Halos}
\author{Yu-Ming Yang}
\email{yangyuming@ihep.ac.cn}
\affiliation{State Key Laboratory of Particle Astrophysics, Institute of High Energy Physics, Chinese Academy of Sciences, Beijing 100049, China}
\affiliation{School of Physical Sciences, University of Chinese Academy of Sciences, Beijing 100049, China }

\author{Xiao-Jun Bi}
\email{bixj@ihep.ac.cn}
\affiliation{State Key Laboratory of Particle Astrophysics, Institute of High Energy Physics, Chinese Academy of Sciences, Beijing 100049, China}
\affiliation{School of Physical Sciences, University of Chinese Academy of Sciences, Beijing 100049, China }

\author{Long Wang}
\email{wanglong8@sysu.edu.cn}
\affiliation{School of Physics and Astronomy, Sun Yat-sen University, Zhuhai 519082, China}

\author{Peng-Fei Yin}
\email{yinpf@ihep.ac.cn}
\affiliation{State Key Laboratory of Particle Astrophysics, Institute of High Energy Physics, Chinese Academy of Sciences, Beijing 100049, China}

\begin{abstract}
Interference of the ultralight dark matter (ULDM) field generates time-varying gravitational potential fluctuations, which stochastically heat stellar systems embedded in ULDM halos. Small-sized stellar systems are therefore often used to set stringent constraints on ULDM.
However, the evolution of systems with sizes well below the ULDM de Broglie wavelength remains poorly explored.
Using numerical simulations, we show that the evolution of compact stellar systems in ULDM halos is governed by the interplay between internal stellar relaxation and ULDM-induced heating. We find the following main results.
First, in sufficiently compact systems, relaxation-driven core collapse dominates, allowing the system to remain bound and dense, while ULDM-induced stripping of outer stars further accelerates the collapse. 
Second, in more extended systems, ULDM heating dominates and ultimately disrupts the system. Near the disruption threshold, we identify systems resembling ultra-faint dwarfs like Segue 1. 
Third, we further introduce a dimensionless parameter to quantify the relative importance of heating and relaxation and finally lead to an evolutionary phase diagram. Our results reveal the rich and nontrivial dynamics of compact stellar systems in ULDM halos, indicating that precise system modeling is essential for robust ULDM constraints.
\end{abstract}

\maketitle
\noindent \textit{\textbf{Introduction.}}---The interference of ultralight dark matter (ULDM)  \cite{Hu_2000, Peebles_2000, Hui_2017, Hui_2021} waves produces stochastic density granules in ULDM halos. These inhomogeneities have been proposed as a possible explanation for several gravitational lensing anomalies \cite{Amruth:2023xqj, Chan_2020}. Moreover, the associated fluctuations in the gravitational potential induce dynamical heating of stellar systems embedded in ULDM halos \cite{Bar_Or_2019, El_Zant_2019, Church_2019, Chiang:2022rlx, Dutta_Chowdhury_2023, Teodori:2025rul, zhao2025semianalyticmodeleffectsfuzzy, Yang:2024vgw, Yang:2024ixt, Yang:2025bae}, leading to an expansion of their sizes. Consequently, systems with small spatial extents, such as
nuclear star clusters \cite{Marsh:2018zyw, Schive:2019rrw, Chiang:2021uvt} and ultra-faint dwarf galaxies (UFDs) \cite{Dalal:2022rmp, May:2025ppj}, are often regarded
as imposing strong lower bounds on the ULDM particle mass. However, a recent study \cite{Eberhardt:2025lbx} suggests that for systems much smaller than the ULDM de Broglie wavelength, such heating may remain ineffective over a long time.

Unlike larger systems, the relaxation time of compact systems can be shorter than the age of the Universe \cite{2008gady.book.....B, Meylan_1997}. As a result, effects driven by two-body gravitational scattering, such as core collapse and envelope expansion, may play a prominent role in their dynamical evolution \cite{2008gady.book.....B}. While these effects are well established in star cluster simulations \cite{Meylan_1997, Wang_2016}, the evolutionary behavior of such systems embedded within ULDM halos remains poorly explored.  Investigating this issue through reliable numerical simulations presents significant computational challenges, as it demands high temporal and spatial resolution.

In this study, we present a simulation framework that integrates ULDM simulations \cite{Edwards_2018} with direct $N$-body modeling of stellar systems \cite{Wang_2020}. 
We employ PeTar \cite{Wang_2020}, a widely used code for star cluster simulations, to simulate the internal evolution of stellar systems under stochastic heating
induced by ULDM. Our results demonstrate that the interplay between stellar relaxation and ULDM heating governs the evolution of stellar systems. For sufficiently dense systems, we find that they can remain bound and compact throughout their evolution, even for an ULDM particle mass $m_a$ as low as $10^{-22}$ eV. Moreover, the stripping of outer stars driven by ULDM heating could accelerate core collapse and yield systems even more concentrated than their initial states. In contrast, initially less compact systems are more vulnerable to disruption. In near-disruption cases, we identify remnants whose properties resemble the observed UFDs, such as Segue 1 \cite{Belokurov_2007, Martin_2008, Simon_2011, Simon_2019}. Finally, we introduce a dimensionless parameter that quantifies the relative strength of ULDM heating versus stellar relaxation. Through extensive simulations, we construct a two-dimensional phase diagram that captures the distinct evolutionary fates of stellar systems in ULDM halos.


\noindent \textit{\textbf{Simulation setup.}}---Since the spatial extent of the stellar system ($\lesssim 30$ pc) under study is much smaller than the ULDM de Broglie wavelength ($\sim $ kpc), resolving its evolution demands much finer temporal and spatial resolution than the ULDM simulation itself. To reduce computational cost, we adopt a two-stage approach: first, we model the entire stellar system as a mass point evolving within the ULDM halo, recording its trajectory and the associated ULDM-induced tidal tensor; subsequently, we use a $N$-body code to simulate the stellar system’s internal evolution under this time-dependent tidal tensor.

In the first-stage simulation, we account for tidal stripping of the ULDM halo by the Milky Way (MW) and the Large Magellanic Cloud (LMC), adopting a realistic assembly history where the halo evolves outside the MW for $\sim 3$ Gyr before infalling onto an MW orbit similar to Segue 1's inferred trajectory \cite{Pace_2022}. This setup enables a meaningful comparison with observations. To ensure a fair comparison among second-stage simulations with different initial internal stellar configurations, we only perform a single first-stage simulation for $m_a=10^{-22}\mathrm{\, eV}$, where the stellar system is modeled as a point mass of $M_\star=10^3M_\odot$ with a softening length of 5 pc. All second-stage simulations adopt the same tidal tensor obtained from this first-stage run. This simplification does not qualitatively affect our conclusions. Further details of the first-stage simulation, including the construction of the initial condition, the implementation of MW and LMC tidal forces, and the system evolution, are provided in the Supplemental Material.

The impact of ULDM on the internal evolution of the stellar system originates from the inhomogeneity of its gravitational potential $V_\mathrm{ULDM}$ \cite{footnote1}.  Since the characteristic variation scale of $V_\mathrm{ULDM}$ is much larger than the size of the stellar system, the ULDM-induced acceleration of a star at position $\boldsymbol{r}$ with respect to the system’s center of mass $\boldsymbol{r}_c$ can be approximated by 
\begin{equation}
    \Delta\boldsymbol{a}(\boldsymbol{r},\boldsymbol{r}_c)\simeq -(\boldsymbol{r}-\boldsymbol{r}_c)\cdot\nabla\nabla V_\text{ULDM}(\boldsymbol{r}_c),
    \label{tidal_tensor}
\end{equation}
where $\nabla\nabla V_\text{ULDM}$ denotes the tidal tensor obtained from the first-stage simulation, and $\boldsymbol{r}_c$ is the center-of-mass position of the stellar system within the ULDM halo.

To illustrate the diversity of evolutionary outcomes, we examine six representative simulations. The initial stellar density distribution is assumed to follow a Plummer profile \cite{1911MNRAS..71..460P} $\rho_\star(r)=(3M_\star/4\pi R_\mathrm{h}^3)(1+r^2/R_\mathrm{h}^2)^{-5/2}$, with initial conditions generated in approximate dynamical equilibrium using McLuster \cite{K_pper_2011}. The adopted total stellar mass $M_\star$, particle number $N$, and two-dimensional half-mass radius $R_\mathrm{h}$ are listed in Tab. \ref{Tab1}.

In the second-stage simulation, we evolve the stellar particles using the direct $N$-body code PeTar \cite{Wang_2020}. For each parameter set, we perform an ULDM-influenced run and an isolated run without ULDM. By the end of simulations ($13$ Gyr), the accumulated energy error in all six isolated runs remains below $0.015\%$. In the simulations with ULDM, the tidal tensor is incorporated into PeTar by fixing the system’s center of mass at $\boldsymbol{r}_c=\boldsymbol{0}$ and computing the additional acceleration induced by ULDM for each stellar particle according to Eq. (\ref{tidal_tensor}).

\begin{figure}[htbp]
    \centering
    \includegraphics[width=\linewidth]{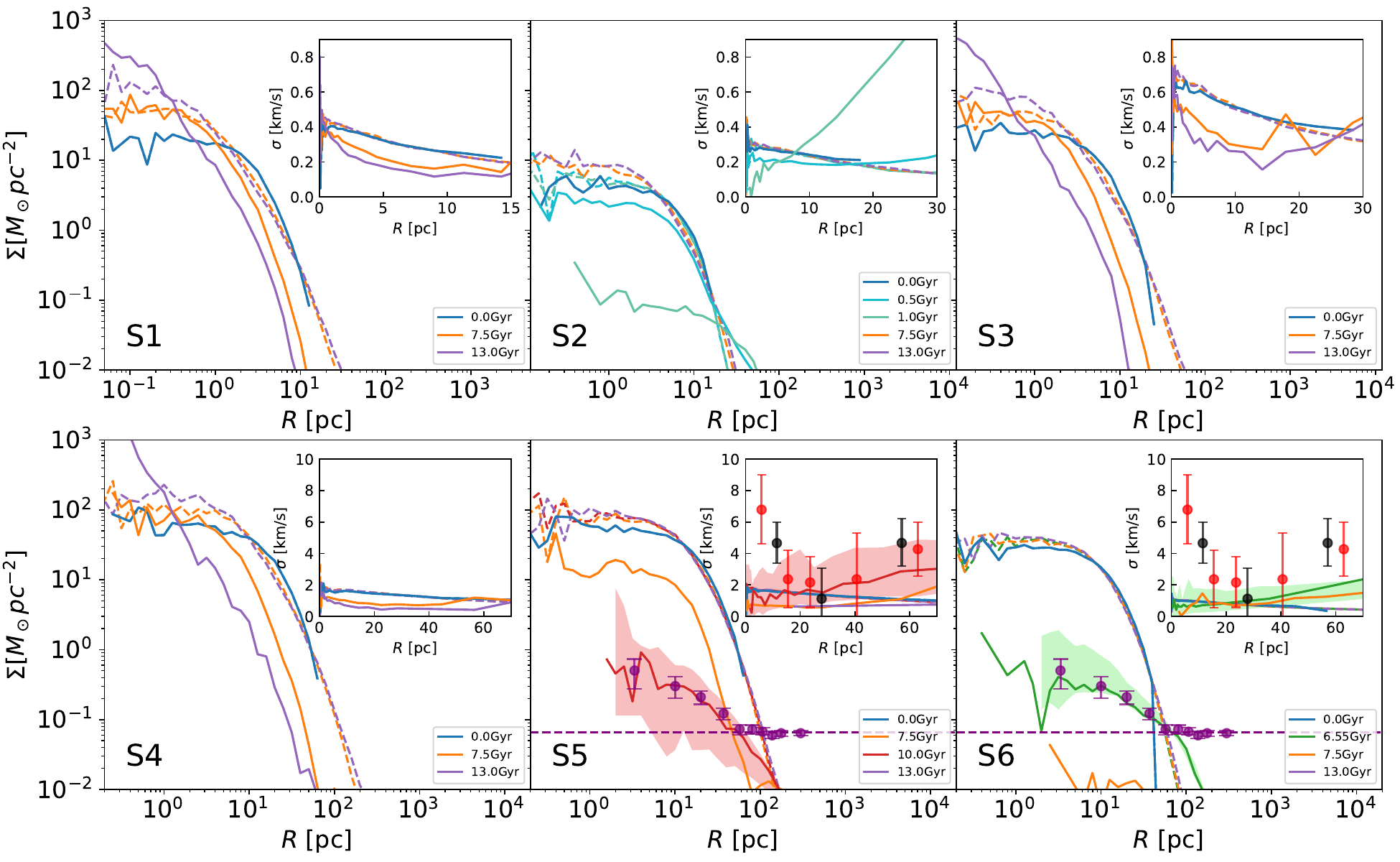}
    \caption{Circularly averaged stellar surface density profiles of six simulation systems at different snapshots, obtained by projecting onto the $z=0$ plane and azimuthal averaging. Insets show the corresponding velocity dispersion profiles along the $z$-direction. Colors indicate evolutionary time, while solid and dashed lines denote evolution with and without ULDM, respectively. In the last two panels, the light-red and light-green shaded regions represent fluctuations arising from 100 randomly selected viewing directions. Observational data for Segue~1 are also shown, including the stellar surface density (purple points) \cite{Martin_2008} and line-of-sight velocity dispersion (red and black) profiles \cite{Simon_2011}.}
    \label{density_velocity}
\end{figure}

Fig. \ref{density_velocity} shows the temporal evolution of both the circularly averaged projected stellar density and the velocity dispersion along the $z$-direction for the six stellar systems initialized with parameters from Tab. \ref{Tab1}. Solid and dashed lines represent systems evolving within an ULDM halo and in isolation, respectively, with distinct colors denoting different evolutionary times.



\begin{table}[htbp]
    \centering
    \caption{Initial condition parameters of the stellar systems.}
    \begin{ruledtabular}
    \begin{tabular}{ccccccc}
    &S1&S2&S3&S4&S5&S6\\
    \hline
    $M_\star[10^3 M_\odot]$&$1$&$1$&$5$&$100$&$100$&$25$ \\
    $N[10^4]$&$1$&$1$&$1$&$1$&$3$&$5$ \\
    $R_\mathrm{h}[\mathrm{pc}]$&$5$&$10$&$10$&$30$&$30$&$20$ \\
    \end{tabular}
    \end{ruledtabular}
    \label{Tab1}
\end{table}

\begin{figure}[htbp]
    \centering
    \includegraphics[width=0.9\linewidth]{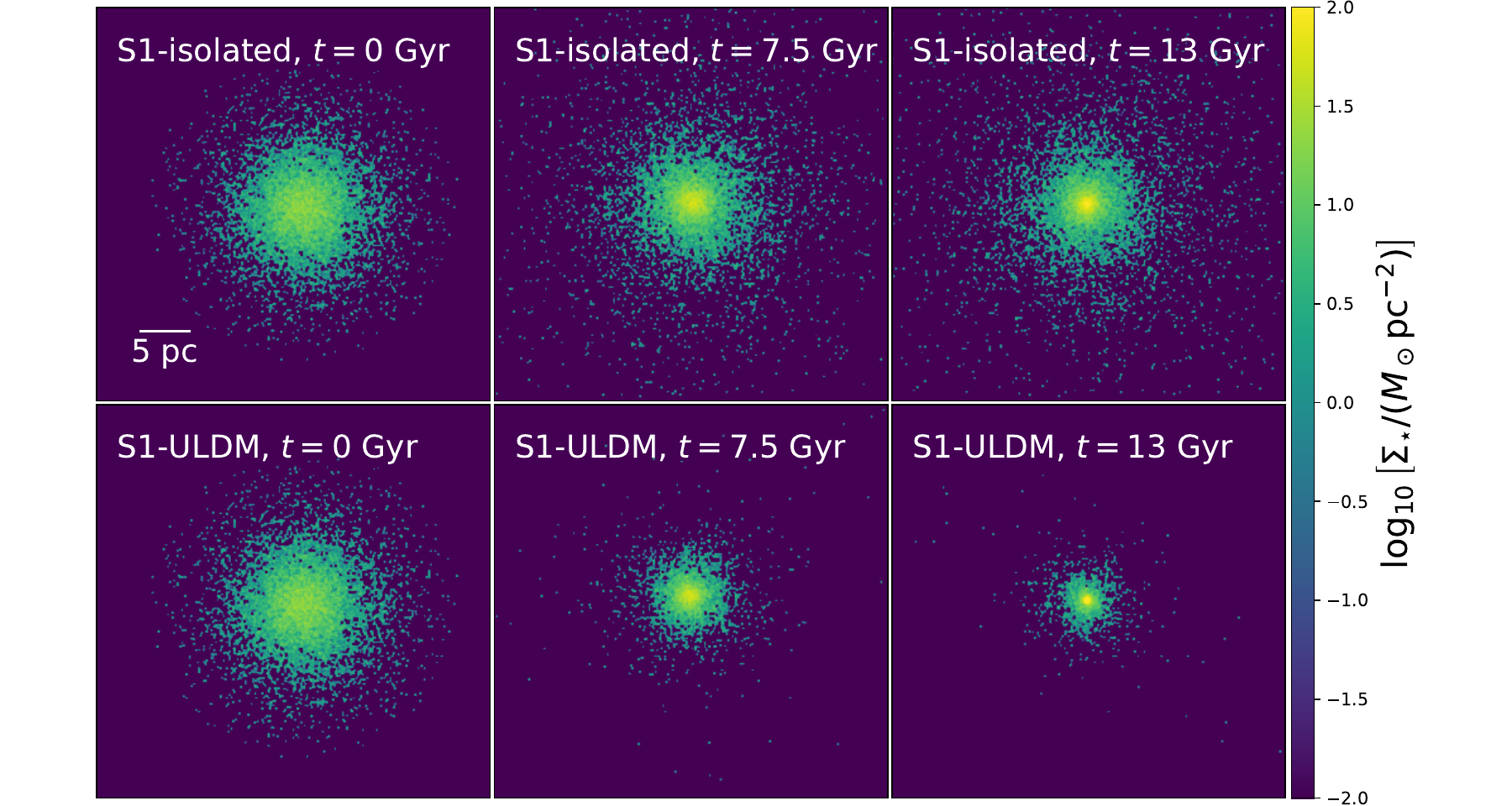}
    \caption{Stellar surface density maps of S1 at different simulation times, obtained by projecting stellar particles onto the $z=0$ plane and constructing the density using a particle-mesh method \cite{Vogelsberger:2019ynw}. The upper and lower panels show evolution without and with ULDM, respectively.}
    \label{stellar_surface}
\end{figure}

\begin{figure}[htbp]
    \centering
    \includegraphics[width=0.8\linewidth]{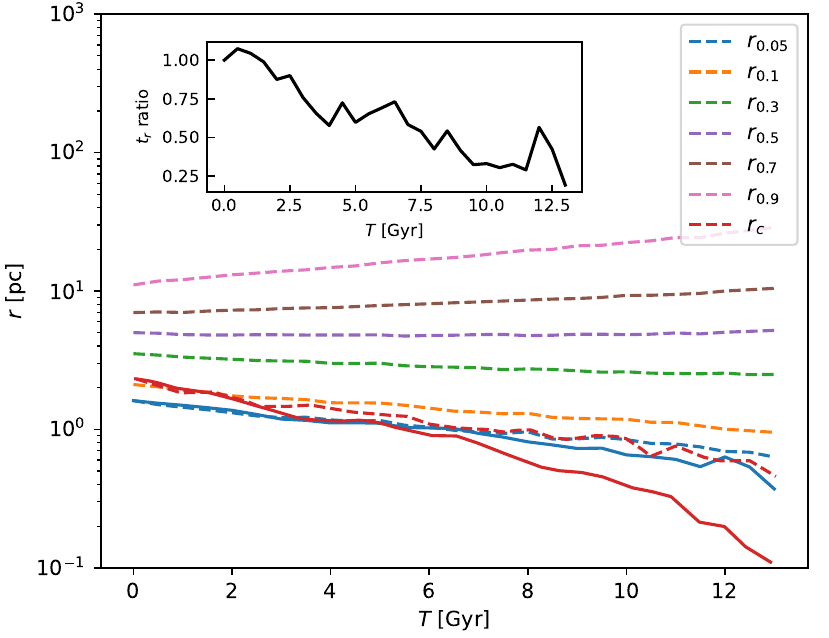}
    \caption{Time evolution of the core radius $r_c$ and Lagrangian radii of S1, with colors indicating different enclosed mass fractions. Solid and dashed lines denote evolution with and without ULDM, respectively. The inset shows the ratio of the averaged relaxation time within $r_{0.05}$ for the cases with and without ULDM.}
    \label{relaxation_time}
\end{figure}

\noindent \textit{\textbf{Stellar dynamics without ULDM.}}--- 
In the isolated evolution of S1, core collapse is evident from the rising central stellar density and the concurrent contraction of the core radius, as shown in the first panel of Fig. \ref{density_velocity}.
This is further illustrated in the upper panels of Fig. \ref{stellar_surface}, where the stellar surface density projected onto the $z=0$ reveals
the enhancement of the central density. Fig. \ref{relaxation_time} shows the evolution of the core radius (defined as \cite{1985ApJ...298...80C}) and Lagrangian radii \cite{lagrangian}, indicating both progressive core contraction and outer envelope expansion (e.g., increasing $r_{0.9}$) as energy is transported outward during core collapse.

Core collapse is driven by two-body gravitational scattering among stars, with a characteristic timescale that scales with the relaxation time \cite{1987degc.book.....S}, $t_\mathrm{r}=0.065\langle v^2\rangle^{3/2}/(\rho\langle m\rangle G^2\ln{\Lambda})$, where $\langle m \rangle$ is the mean stellar mass, $\ln \Lambda \sim \ln N$ is the Coulomb logarithm roughly determined by $N$, and the mean-squared stellar velocity $\langle v^2\rangle$ and density $\rho$ are position dependent quantities. This expression describes the relaxation time evaluated locally. For the global system evolution, the relevant timescale is the half-mass relaxation time \cite{1987degc.book.....S, Meylan_1997},
\begin{equation}
    t_\mathrm{rh}=0.138\frac{{M_\star}^{1/2}r_\mathrm{h}^{3/2}}{\langle m\rangle G^{1/2}\ln\Lambda}\simeq \frac{0.212N R_\mathrm{h}^{3/2}}{{M_\star}^{1/2} G^{1/2}\ln\Lambda},
    \label{t_rh}
\end{equation}
where $r_\mathrm{h}$ is the three-dimensional half-mass radius, taken to be $r_\mathrm{h}\simeq(4/3)R_\mathrm{h}$ under the assumption of spherical symmetry \cite{Simon_2019}.

The parameter dependences implied by these equations are consistent with the isolated evolution results (dashed lines in Fig.~\ref{density_velocity}). For example, S2 differs from S1 only by a larger initial size $R_\mathrm{h}$ and therefore exhibits a slower core collapse. By contrast, S3 has a higher total stellar mass $M_\star$ than S2 and undergoes a more pronounced collapse, while the larger particle number $N$ in S5 compared to S4 leads to a slower evolution.

\noindent \textit{\textbf{Accelerated core collapse.}}---Core collapse tends to make the system more centrally concentrated, whereas ULDM heating drives it toward a more diffuse state. Intuitively, these two effects are expected to compete during evolution, with their relative importance depending on initial conditions, leading to distinct evolutionary regimes.

Remarkably, our simulations reveal that in  core collapse-dominated regimes (exemplified by systems S1, S3, and S4), ULDM heating not only fails to counteract collapse but actually accelerates it. This is clearly evidenced by comparing the ULDM-influenced (solid lines) and isolated (dashed lines) evolutionary results in the relevant panels of Fig. \ref{density_velocity}. To further highlight this phenomenon, the lower panels of Fig. \ref{stellar_surface} show the ULDM-influenced evolution of the stellar surface density of S1. It is evident that ULDM heating strips stars from the outer regions, leading to a central stellar distribution that is more compact than in isolated evolution by 13 Gyr. Consistently, a comparison of the red solid and dashed lines in Fig. \ref{relaxation_time} shows that the contraction of the core radius proceeds significantly faster within an ULDM halo. 


This acceleration of core collapse can be understood from  the first equality in Eq.~\ref{t_rh}: ULDM heating reduces both the bound stellar mass and the half-mass radius, thereby shortening the relaxation timescale. Alternatively, heating-induced mass loss lowers the stellar gravitational potential and hence reduces the velocity dispersion $\sigma$ (inset of the first panel of Fig.~\ref{density_velocity}). Through the dependence of the local relaxation time $t_\mathrm{r}$ on $v^2\simeq3\sigma^2$, this reduction naturally shortens the relaxation time and enhances two-body relaxation. As shown in the inset of Fig.~\ref{relaxation_time}, the averaged relaxation time within $r_{0.05}$ drops to $\sim 25\%$ of that in the isolated case by the end of the evolution. From a thermodynamic perspective, the ULDM-induced steepening of the inner velocity dispersion gradient (inset of the first panel of Fig.~\ref{density_velocity}) amplifies the temperature contrast between the inner and outer regions, driving more efficient outward heat transport and naturally explaining the accelerated core collapse. Interestingly, a similar phenomenon has been reported by studies of core collapse in self-interacting dark matter halos \cite{Nishikawa_2020}.

\noindent \textit{\textbf{Disruption of stellar system.}}---We find that ULDM-induced heating becomes increasingly important as the initial stellar distribution becomes more diffuse. Beyond a critical threshold, this heating ultimately leads to the disruption of the stellar system. 

This trend is evident from a comparison of S1 and S2: both have the same total stellar mass and particle number, but S2 has a larger initial half-mass radius of 10 pc. As shown in the second panel of Fig. \ref{density_velocity}, ULDM heating nearly disperses the stellar distribution by $\sim 1$ Gyr. Meanwhile, the velocity dispersion increases rapidly as energy is continuously injected into the stellar system, driving it from a self-bound configuration toward a dispersing state. This outcome arises in part because the increased half-mass radius lowers the initial stellar density, thereby reducing the dominance of the stellar gravitational potential relative to that of ULDM. In addition, the relaxation process is weakened owing to the scaling $t_\mathrm{rh}\propto R_\mathrm{h}^{3/2}$, causing the heating effect to dominate over relaxation. However, even with a relatively large half-mass radius, increasing the particle mass by a factor of five causes the system S3 to become relaxation-dominated once again, as shown in the third panel of Fig. \ref{density_velocity}.

The systems S4 and S5 have identical initial density distributions but differ in particle number, so the smooth stellar potential is the same while local two-body relaxation is stronger in S4. Consequently, S4 undergoes core collapse, with ULDM heating further enhancing the process and leaving the system stable. In contrast, weaker relaxation in S5 allows heating to dominate, driving the system near disruption by $\sim 10$ Gyr. This comparison intuitively demonstrates that, under otherwise identical conditions, variations in local relaxation strength can produce qualitatively distinct evolutionary outcomes.

\noindent \textit{\textbf{Near disruption: Segue 1-analog system.}}---To facilitate comparison with observations, we take a UFD Segue 1 as an example, presenting its observed stellar surface density and line-of-sight velocity dispersion profiles in the last two panels of Fig. \ref{density_velocity}. This stellar density profile is derived from a sample of fewer than 100 stars \cite{Martin_2008}. After subtracting the background, we normalize the total stellar mass to $\sim 800 M_\odot$, within the observationally inferred range given uncertainties in the mass-to-light ratio and other systematics. The velocity dispersion data are adopted from \cite{Simon_2011}, with red and black points representing measurements obtained by grouping 15 and 23 stars per bin, respectively.

When the system S5 approaches disruption at $t=10$ Gyr (red solid line), the derived stellar surface density and velocity dispersion profiles broadly match observations, except that the sharply rising central velocity dispersion in the data is not reproduced. This rise may reflect additional physical mechanisms such as a central supermassive black hole \cite{Lujan:2025auq}. The light-red shaded region represents fluctuations arising from 100 randomly selected viewing directions. Owing to the limited number of remaining particles within 60 pc ($\sim 400$) at this near-disruption stage, projections along different lines of sight exhibit large variations, which broadly encompass the observed data. 

Note that S5's stellar particles mass $\sim3.3 M_\odot$ is not representative of a realistic stellar mass spectrum \cite{Kroupa_2002}. We also find that Segue 1-analog systems can be reproduced with more realistic stellar masses. For example, in S6, where the individual stellar particle mass is $0.5 M_\odot$, the stellar surface density and velocity dispersion profiles at $t\simeq6.55\mathrm{\, Gyr}$ also achieve good agreement with those of Segue 1 (last panel of Fig.~\ref{density_velocity}), albeit at a younger age than the inferred $\sim13\mathrm{\, Gyr}$. 

Overall, our results suggest that for an ULDM particle mass of $10^{-22}\mathrm{\,eV}$, Segue~1 may be interpreted as a system near disruption \cite{Eberhardt:2025lbx}. We emphasize that the goal of this study is to illustrate a plausible evolution scenario for compact stellar systems. More generally, their evolution within ULDM halos reflects a complex interplay between two-body relaxation and ULDM-induced heating.  Incorporating additional physics, such as a stellar initial mass function \cite{Kroupa_2002} and stellar evolution, will further complicate this picture, indicating that robust constraints on ULDM require more detailed modeling. 

\noindent \textit{\textbf{Phase diagram.}}---Since core collapse is driven by two-body relaxation, its timescale scales with the relaxation time, $t_\mathrm{rh}$. By contrast, in the regime $R_\mathrm{h} \ll \lambda_\mathrm{dB}$, we adopt the phenomenological estimate proposed in Ref.~\cite{Eberhardt:2025lbx} as a characteristic timescale for ULDM-driven disruption, $t_\mathrm{dis}=(3M_\star/4\pi\bar{\rho}_\mathrm{ULDM}R^3_\mathrm{h})(\lambda_\mathrm{dB}/\sigma_\mathrm{ULDM})$, where $\lambda_\mathrm{dB}$ and $\sigma_\mathrm{ULDM}$ are the ULDM de Broglie wavelength and velocity dispersion, respectively. As shown in the Supplemental Material, the stellar system orbits within $r_\mathrm{orbit}\simeq 3 \mathrm{\, kpc}$ of the halo center. Therefore, we take $\bar{\rho}_\mathrm{ULDM}=3M(<r_\mathrm{orbit})/4\pi r^3_\mathrm{orbit}$ to be the mean ULDM density within this radius. We emphasize that $t_\mathrm{rh}$ and $t_\mathrm{dis}$ should be interpreted as characteristic parameters rather than precise evolutionary times, which may differ by factors of $\mathcal{O}(1)$. 

The competition between heating and relaxation can be quantified by the ratio
\begin{equation}
\begin{aligned}
    &\alpha\equiv\frac{t_\mathrm{rh}}{t_\mathrm{dis}}=0.153\left(\frac{N}{10^4}\frac{4}{\lg N}\right)\left(\frac{10^3M_\odot}{M_\star}\right)^{3/2}\left(\frac{R_\mathrm{h}}{10\mathrm{\,pc}}\right)^{9/2}\\
    &\times\left(\frac{m_ac^2}{10^{-22}\mathrm{\,eV}}\right)\left(\frac{M(<r_\mathrm{orbit})}{3.6\times 10^8M_\odot}\right)\left(\frac{3\mathrm{\,kpc}}{r_\mathrm{orbit}}\right)^3\left(\frac{\sigma_\mathrm{ULDM}}{20\mathrm{\,km\,s}^{-1}}\right)^2,
\end{aligned}
\label{t_rh_t_dis_ratio}
\end{equation}
with larger $\alpha$ indicating increasing dominance of ULDM heating. Based on extensive simulations covering a large stellar system parameter space,
we find that for $t_\mathrm{rh} \gtrsim 4.7 \mathrm{\, Gyr}$ ($t_\mathrm{dis} \gtrsim 85.0 \mathrm{\, Gyr}$), relaxation (heating) has a negligible impact over the age of the Universe ($\sim 13 \mathrm{\, Gyr}$). Meanwhile, we identify a threshold $\alpha = 0.05$, above (below) which heating (relaxation) dominates the stellar system’s evolution for moderate $N$. 

\begin{figure}[htbp]
    \centering
    \includegraphics[width=\linewidth]{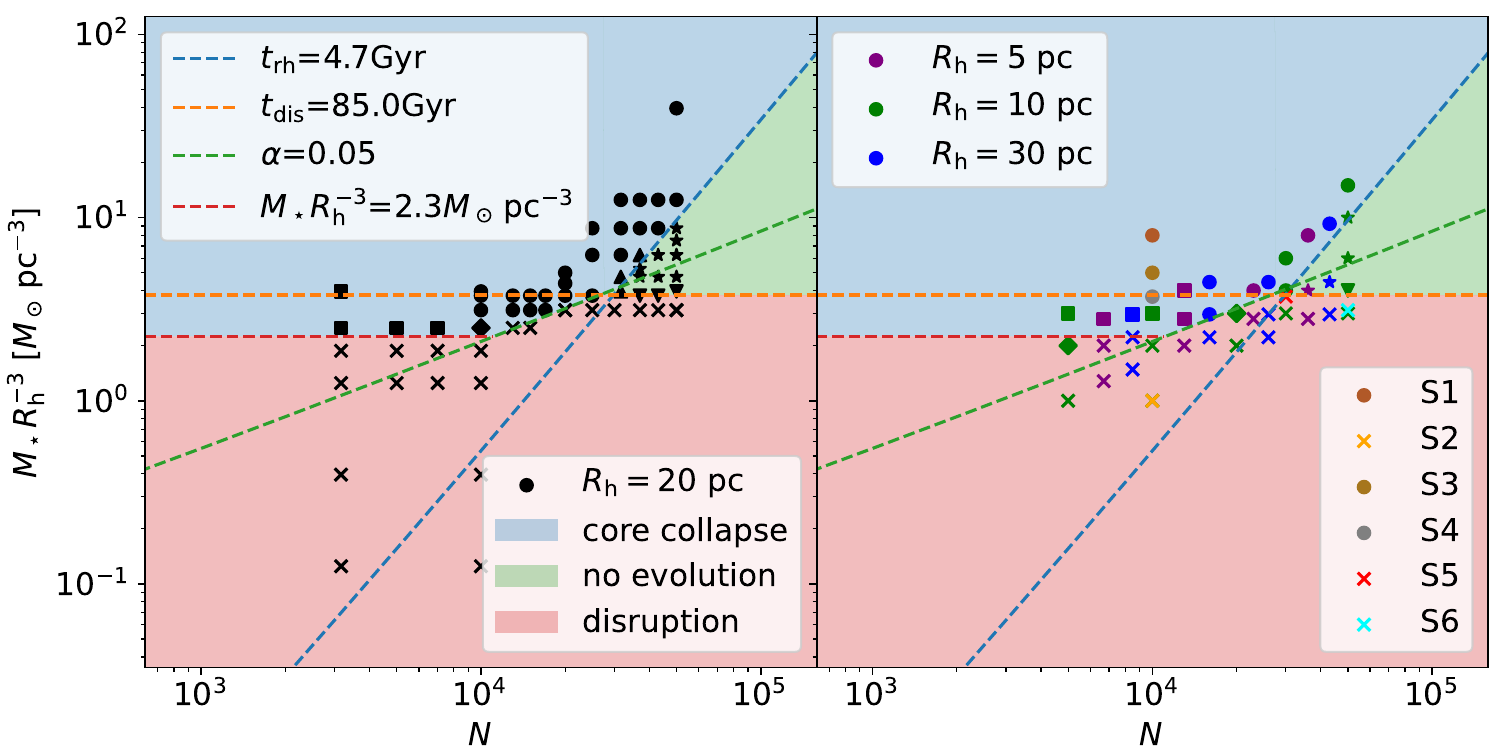}
    \caption{Phase diagram of stellar systems within ULDM halos for $m_a=10^{-22}$eV. Filled circles, crosses, and pentagrams denote systems undergoing core collapse, disruption, and negligible evolution, respectively. Squares mark systems with strong collapse that are susceptible to numerical artifacts. Upward triangles, downward triangles, and diamonds indicate intermediate cases between collapse and no evolution, disruption and no evolution, and collapse and disruption, respectively, and are therefore difficult to classify unambiguously. Colors encode different $R_\mathrm{h}$. The six simulation sets S1-S6 are indicated separately.} 
    \label{t_rh_t_dis}
\end{figure}

We find that 
stellar systems with different initial parameters $(M_\star, N, R_\mathrm{h})$ can be mapped onto a single $(N, M_\star R_\mathrm{h}^{-3})$ plane (Fig.~\ref{t_rh_t_dis}), which separates into three regions with distinct evolutionary outcomes. This reduction of parameter space reflects the degeneracy between $M_\star$ and $R_\mathrm{h}$, consistent with the analytic scalings of $t_\mathrm{rh}$, $t_\mathrm{dis}$, and $\alpha$. Moreover, the boundaries between different regions have clear physical interpretations. 

As shown in Fig. \ref{t_rh_t_dis}, the blue, orange, and green dashed lines correspond to loci of parameter points satisfying the three critical values introduced above: $t_\mathrm{rh} = 4.7\mathrm{\, Gyr}$, $t_\mathrm{dis} = 85.0\mathrm{\, Gyr}$, and $\alpha = 0.05$, respectively. The evolutionary behavior in each region can be summarized as follows. In the light-green region ($t_\mathrm{rh}>4.7\mathrm{\, Gyr}$ and $t_\mathrm{dis}>85.0\mathrm{\, Gyr}$), both relaxation and heating are ineffective over 13 Gyr, and the systems remain quite stable. In the light-blue region ($\alpha<0.05$ and $t_\mathrm{rh}<4.7 \mathrm{\, Gyr}$), relaxation dominates, leading to core collapse. By contrast, systems in the light-red region, largely satisfying $\alpha>0.05$ and $t_\mathrm{dis}<85.0\mathrm{\, Gyr}$, evolve toward disruption under dominant ULDM heating. Nevertheless, a small subset of the light-red region falls at $\alpha<0.05$, because the stellar density there is sufficiently low that, even if partial core collapse occurs, the system is ultimately dispersed by ULDM heating. As shown in Fig.~\ref{t_rh_t_dis}, the different simulated evolutionary outcomes, denoted by distinct symbols, occupy well-defined and separated regions of parameter space.

Another feature revealed by Fig. \ref{t_rh_t_dis} is that the lines defined by the three critical values intersect at nearly a single point, even though each line is determined independently from the simulation results. This is not a coincidence and can be understood as follows. At the intersection of the $t_\mathrm{rh}$ and $t_\mathrm{dis}$ boundaries, both relaxation and heating become relevant at $\sim 13 \mathrm{\,Gyr}$. As a result, the two effects are comparable at this point, and it therefore naturally lies on the critical $\alpha$ boundary. This further demonstrates the consistency between the analytic scalings and the simulations.

Finally, we emphasize that the above analysis is performed at fixed ULDM particle mass, halo, and orbital parameters, and therefore probes only the dependence on the initial stellar system parameters. Whether the scaling with these former quantities follows Eq. \ref{t_rh_t_dis_ratio} remains to be tested.

\noindent \textit{\textbf{Conclusions.}}---In summary, we demonstrate how compact stellar systems evolve under the interplay between internal stellar relaxation and ULDM-induced heating.
We identify distinct evolutionary regimes and present a phase diagram. Although this diagram is constructed by considering only the intrinsic parameters of the stellar system, it already reveals the rich and nontrivial nature of the evolution.
Future works should extend this analysis to explore the dependence on ULDM particle and halo properties. Moreover, incorporating more realistic ingredients, such as a stellar initial mass function and stellar evolution, will also be essential for establishing more robust connections with current and forthcoming observations.


\noindent \textit{Acknowledgements.}---We thank Shuo Li and Hui Li for useful discussions. This work is supported by the National Natural Science Foundation of China under grants No.12447105, No.12575113, No.12573041, and No.12233013. L.W. thanks the High-level Youth Talent Project (Provincial Financial Allocation) through the grant 2023HYSPT0706, the Fundamental Research Funds for the Central Universities, Sun Yat-sen University (2025QNPY04).

\bibliography{Refs}
\bibliographystyle{apsrev4-1}

\onecolumngrid
\setcounter{equation}{0}
\setcounter{figure}{0}
\newpage
\begin{center}
	\textbf{\large Supplemental Material for Collapse versus Disruption: The Fate of Compact Stellar Systems in Ultralight Dark Matter Halos} \\ 
	\vspace{0.05in}
	{Yu-Ming Yang$^{1,2}$, Xiao-Jun Bi$^{1,2}$, Long Wang$^{3}$ and Peng-Fei Yin$^1$}
	\vspace{0.05in}
\end{center}
\centerline{{\it  $^1$State Key Laboratory of Particle Astrophysics, Institute of High Energy Physics,}}
\centerline{{\it Chinese Academy of Sciences, Beijing 100049, China}}
\centerline{{\it  $^2$School of Physical Sciences, University of Chinese Academy of Sciences, Beijing 100049, China}}
\centerline{{\it $^3$School of Physics and Astronomy, Sun Yat-sen University, Zhuhai 519082, China}}
\vspace{0.05in}
\setcounter{page}{1}

This Supplemental Material describes the first-stage simulation, including the construction of the initial condition, the implementation of the Milky Way (MW) and Large Magellanic Cloud (LMC) tidal fields, and the evolution of the stellar system treated as a point mass within the ultralight dark matter (ULDM) halo.
\twocolumngrid
\noindent \textit{\textbf{Initial condition construction.}}--We construct the initial ULDM halo wavefunction based on an input soliton-Navarro-Frenk-White profile \cite{Schive:2014dra, Schive:2014hza, Veltmaat:2018dfz, Liao:2024zkj, Mocz:2017wlg, Schwabe:2016rze, Chan:2021bja, Blum:2025aaa}
\begin{equation}
    \rho_\mathrm{in}(r)=\left\{\begin{aligned}
        &\frac{\rho_c}{\left[1+0.091(r/r_c)^2\right]^8},\quad r<kr_c\\
        &\frac{\rho_s}{(r/r_s)\left(1+r/r_s\right)^2},\quad r\geq kr_c,
        \end{aligned}\right.
        \label{input_profile}
\end{equation}
where the solitonic core density is $\rho_c=8.341\times 10^7M_\odot\text{\,kpc}^{-3}$, the core radius $r_c=0.695\text{\,kpc}$, and the matching parameter $k=2.712$. For the outer region, the density and radius parameters are $\rho_s=3.912\times 10^7M_\odot\text{\, kpc}^{-3}$ and $r_s=0.780\text{\, kpc}$, respectively. These parameters are chosen to satisfy both the soliton core density-radius relation $\rho_c=1.95\times 10^7 M_\odot \text{\, kpc}^{-3}\left(m_a/10^{-22}\text{eV}\right)^{-2}\left(r_c/\text{kpc}\right)^{-4}$ and the core-halo mass relation predicted by cosmological ULDM simulations \cite{Schive:2014hza, Veltmaat:2018dfz, Liao:2024zkj}. A detailed description of the parameter selection procedure can be found in \cite{Yang:2025bae}.

We also include the gravitational potential of the stellar system, modeled as a softened point mass, when constructing the initial wavefunction, although its impact is negligible. As described in the main text, we adopt a mass point with $M_0=10^3 M_\odot$ and a softening length of $r_0=5$ pc. The initial wavefunction is constructed using the eigenstate decomposition method \cite{Lin_2018, Yavetz_2022}. Specifically, the wavefunction at $t=0$ is expressed as a linear combination of eigenstates,
\begin{equation}
\psi(0,\boldsymbol{x})=\sum_{nlm}\left|a_{nl}\right|e^{i\phi_{nlm}}\Psi_{nlm}(\boldsymbol{x}),
\end{equation}
where the principal, angular, and magnetic quantum numbers $n,l,$ and $m$ label different eigenstates. The eigenstates $\Psi_{nlm}(\boldsymbol{x})$ are obtained by solving the time-independent Schr$\ddot{\text{o}}$dinger equation,
\begin{equation}
    -\frac{\hbar^2}{2m_a}\nabla^2\Psi_{nlm}+m_a\left[V_\text{in}(r)+V_\star(r)\right]\Psi_{nlm}=E_{nl}\Psi_{nlm},
    \label{eigen_Eq}
\end{equation}
where the spherically symmetric potential $V_\mathrm{in}(r)$ corresponds to the ULDM density profile given in Eq. \ref{input_profile}, and $V_\star(r)=-GM_0/\sqrt{r^2+r_0^2}$ is the softened point-mass potential of the stellar system. 

Owing to the spherical symmetry of the total potential, the eigenstates can be separated as $\Psi_{nlm}(\boldsymbol{x})=R_{nl}(r)Y_l^m(\theta,\phi)$, where $Y_l^m$ are spherical harmonics. We solve the radial wavefunctions $R_{nl}(r)$ using the shooting method. The expansion coefficients $\lvert a_{nl} \rvert$ are then determined via a nonnegative least-squares fit by requiring the random-phase-averaged density profile, $\rho_\mathrm{out}(r)=(m_a/4\pi)\sum_{nl}(2l+1)|a_{nl}|^2R^2_{nl}(r)$, to match the input density profile given in Eq. \ref{input_profile}. The random phases $\phi_{nlm}$ are drawn from a uniform distribution on $[0,2\pi]$. Details of the shooting method, the nonnegative least-squares procedure, and related numerical implementations are provided in \cite{Yang:2024ixt}. The nonzero initial global velocity of the constructed halo is computed using the equations given in \cite{Yang:2024trr} and is subsequently removed via a Galilean boost.

To relax the constructed halo into dynamical equilibrium, we fix the stellar mass point at the origin and evolve the wavefunction in isolation for 1 Gyr. The resulting relaxed state is then adopted as the initial condition for subsequent simulations.

\begin{figure}[htbp]
    \centering    \includegraphics[width=0.45\linewidth]{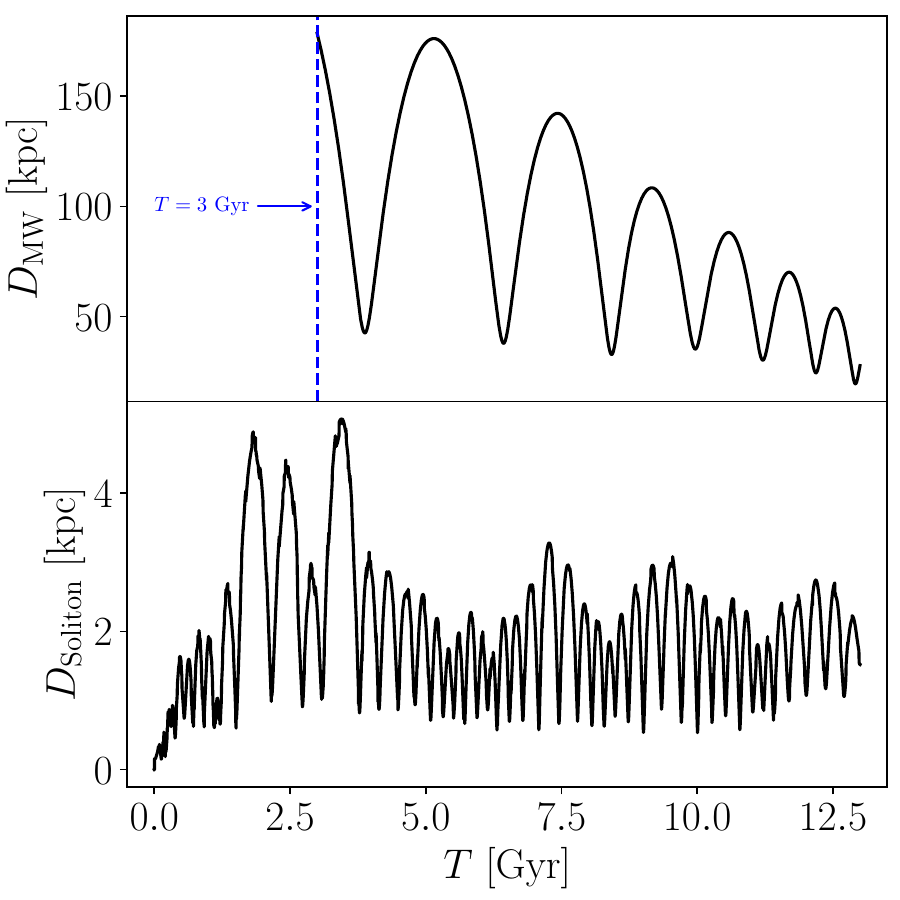}
    \raisebox{0.44cm}{%
        \includegraphics[width=0.492\linewidth]{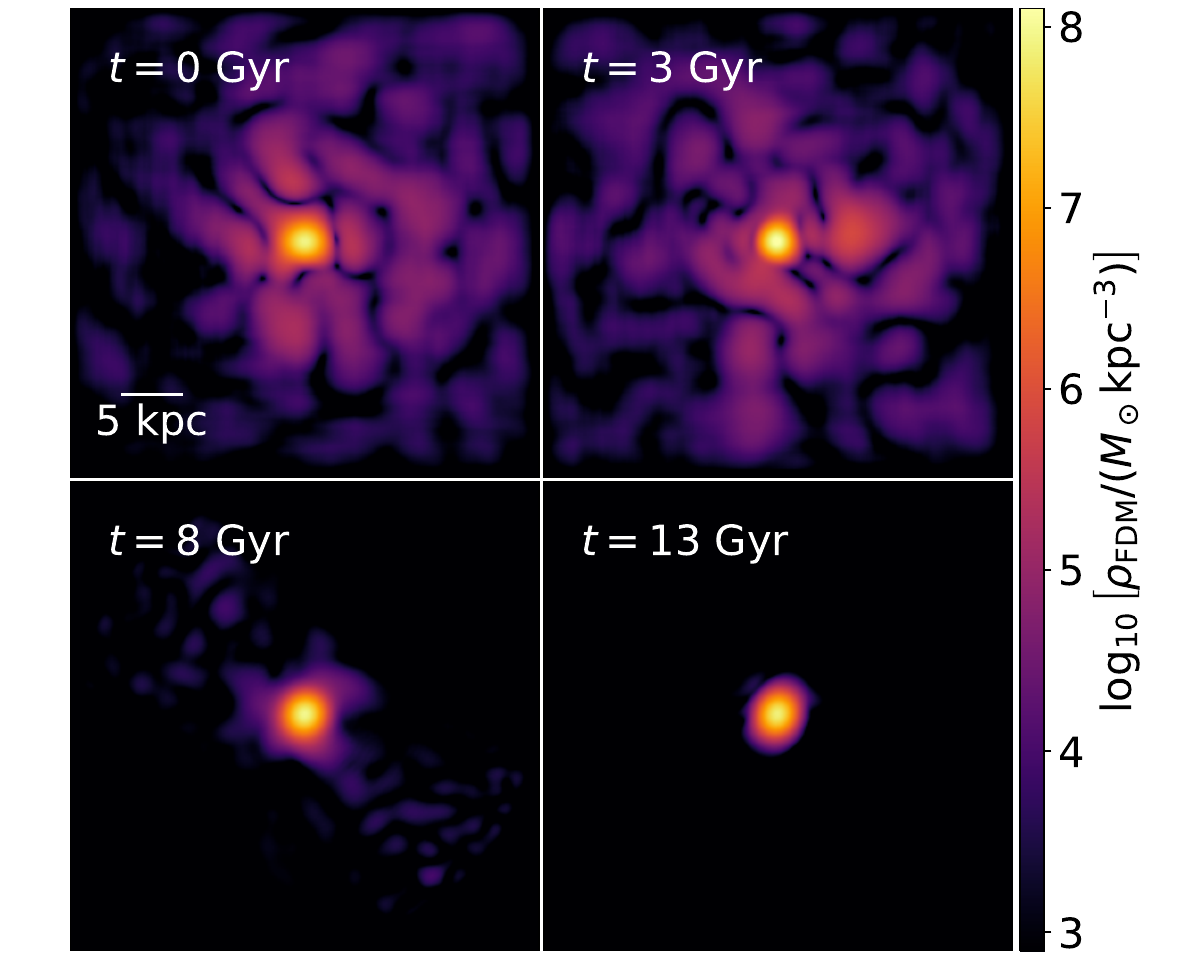}
    }
    \caption{Upper left: orbit of the ULDM halo during infall into the MW following an outside evolution of 3 Gyr. Lower left: temporal evolution of the distance between the stellar system mass point and the soliton center. Right: three-dimensional ULDM density distributions on the $z=0$ plane at four different simulation snapshots.}
    \label{orbits}
\end{figure}

\noindent \textit{\textbf{Tidal potentials of the MW and LMC.}}---As discussed in the main text, to enable meaningful comparison with observations of the ultra-faint dwarf galaxies, we adopt a realistic assembly history in which the ULDM halo evolves outside the MW for 3 Gyr \cite{D_Souza_2021, May:2025ppj}, before falling into the MW on an orbit with a pericenter of $\sim20$ kpc at late times, comparable to that of Segue~1 \cite{Pace_2022}. This orbit is shown in the upper-left panel of Fig.~\ref{orbits}. From $t=3$ Gyr onward, the system is subject to the tidal fields of both the MW and the LMC. The tidal potential induced by either the MW or the LMC can be written in a unified form as \citep{Yang:2025bae}
\begin{equation}
\begin{aligned}
    V^{(I)}_\text{tidal}(\boldsymbol{r}_s,\boldsymbol{r}_I)\simeq &\frac{1}{2}\left[\frac{\partial V_I}{\partial R_I}\frac{(r_{s,x}y_I-r_{s,y}x_I)^2}{(x_I^2+y_I^2)^{3/2}}\right.\\
    &\left.+\frac{\partial^2 V_I}{\partial R_I^2}\frac{(r_{s,x}x_I+r_{s,y}y_I)^2}{x_I^2+y_I^2}+\frac{\partial^2 V_I}{\partial z_I^2}r_{s,z}^2\right.\\
    &\left.+2\frac{\partial^2 V_I}{\partial R_I\partial z_I}\frac{(r_{s,x}x_I+r_{s,y}y_I)r_{s,z}}{(x_I^2+y_I^2)^{1/2}}\right],
\end{aligned}
\label{tidal}
\end{equation}
where the index $I$ denotes either the MW or the LMC. Here, $\boldsymbol{r}_I(t) = (x_I(t), y_I(t), z_I(t))$ is the position of the soliton center relative to the center of the MW or the LMC, and $R_I = (x_I^2 + y_I^2)^{1/2}$ is the corresponding cylindrical radius. The vector $\boldsymbol{r}_s = (r_{s,x}, r_{s,y}, r_{s,z})$ denotes the coordinate relative to the soliton center.

We assume that the soliton center follows the orbit within the MW shown in the upper left panel of Figure \ref{orbits}, which directly determines $\boldsymbol{r}_\mathrm{MW}$. The position $\boldsymbol{r}_\mathrm{LMC}$ is obtained after specifying the orbit of the LMC within the MW; details of this procedure are provided in \cite{Yang:2025bae}. The quantity $V_I$ represents the gravitational potential of the MW or the LMC, expressed in its own central coordinate system, with functional forms and parameters adopted from \cite{Yang:2025bae}. The only difference from \cite{Yang:2025bae} is that we choose the MW dark matter halo density parameter $\rho_{0,h}$ to be the mean value between the weak- and strong-tidal cases listed in their Table 2.

\noindent \textit{\textbf{Evolution of the stellar system as a mass point within the ULDM halo.}}---After constructing the initial ULDM halo wavefunction as described above, the stellar system, modeled as a mass point, is initialized at rest at the origin. The subsequent evolution of the ULDM wavefunction $\psi(t,\boldsymbol{x})$ is governed by the Schr$\ddot{\text{o}}$dinger equation,
\begin{equation}
    i\hbar\frac{\partial\psi}{\partial t}=-\frac{\hbar^2}{2m_a}\nabla^2\psi+m_aV\psi,
    \label{Schrodinger}
\end{equation}
where the ULDM particle mass is fixed to $m_a=10^{-22}$~eV. The total gravitational potential $V$ includes: (1) the ULDM self-gravity potential $V_\mathrm{ULDM}$, obtained by solving the Poisson equation $\nabla^2V_\text{ULDM}=4\pi Gm_a(|\psi|^2-\langle|\psi|^2\rangle)$; (2) the tidal potentials from the MW and LMC, applied after 3 Gyr; and (3) the gravitational potential of the stellar-system mass point. 

Varying the random phases of different eigenstates in the initial wave function while holding all other conditions constant induces distinct interference patterns. Consequently, the orbit of the stellar system within the ULDM halo exhibits stochastic variations. As discussed in the main text, to facilitate a fair comparison of simulation results among stellar systems with different initial internal configurations, we fix the orbits of all stellar systems to be identical in the second-stage simulations. This reference orbit is derived from a single realization of randomly assigned initial phases, modeling the stellar system as a mass point with $10^3 M_\odot$ and a softening length of $5$ pc. This treatment does not qualitatively affect our conclusions, and the quantitative evolutionary differences induced by different orbital configurations will be investigated in future studies.

We use a pseudo-spectral method to evolve the ULDM wave function \cite{Edwards_2018}, adopting a timestep of 1 Myr and a simulation box with a side length of 40 kpc  and a spatial resolution of $256^3$. The stellar system mass point is evolved under the influence of the ULDM potential $V_\text{ULDM}$ according to Newton’s second law, integrated with a fourth-order Runge-Kutta scheme and a timestep of $0.1$ Myr. The temporal evolution of the mass point's distance from the soliton center is presented in the lower left panel of Figure \ref{orbits}. The mass point is initially driven outward by the ULDM heating effect, and eventually settles into oscillations at a distance of approximately $0.8-3.0$ kpc from the soliton center. The three-dimensional ULDM density distributions on the $z=0$ plane at four different snapshots are displayed in the right panel of Figure \ref{orbits}. As shown, after 3 Gyr, the ULDM outside the soliton are gradually stripped away by the tidal forces from the MW and the LMC.

Combining the discussion in the main text with the result shown in the lower-left panel of Fig.~\ref{orbits}, our simulations indicate that the ULDM heating can displace the entire stellar system from the soliton center, at least for compact stellar systems in which the stellar gravitational potential dominates. Such systems are not disrupted and may even undergo core collapse. For larger stellar systems, the influence of core collapse is reduced due to longer relaxation times $t_\mathrm{rh}\propto R_\mathrm{h}^{3/2}$; their evolution is therefore expected to be dominated by ULDM heating. Such systems may not be heated by ULDM as a coherent whole and driven far from the soliton center. Nevertheless, it remains possible that their stellar luminosity peaks are offset from the soliton center. Consequently, the gravitational potential experienced by the stellar system would no longer be spherically symmetric. Previous ULDM constraints derived from spherically symmetric analyses, such as Jeans analyses, may require more careful and refined treatment.

\end{document}